\documentclass[%
 aps,reprint,
showpacs,
 amsmath,amssymb,
 prl,
]{revtex4-1}

\pdfoutput=1
\usepackage{graphicx}
\usepackage{dcolumn}
\usepackage{bm}
\usepackage{wrapfig}
\usepackage{amssymb}
\usepackage{epstopdf}
\usepackage{float}
\usepackage[top=0.75in, bottom=0.75in, left=0.5in, right=0.5in]{geometry}
\usepackage{enumitem}
\setlist[enumerate]{itemsep=0mm}
\begin{document}


\title{Cusps Enable Line Attractors for Neural Computation}
\author{Zhuocheng Xiao$^{1,2}$}
\author{Jiwei Zhang$^{3}$}
\author{Andrew T. Sornborger$^{4,5}$}\email[Corresponding authors' email: ]{sornborg@lanl.gov, taolt@mail.cbi.pku.edu.cn}
\author{Louis Tao$^{2,6,*}$}
\affiliation{$^1$Department of Mathematics, University of Arizona, Tucson, AZ 85721, USA}
\affiliation{$^2$Center for Bioinformatics, National Laboratory of Protein Engineering and Plant Genetic Engineering, School of Life Sciences, Peking University, Beijing, 100871, China}
\affiliation{$^3$Beijing Computational Science Research Center, Beijing, 100193, China}
\affiliation{$^4$CCS-3, Los Alamos National Laboratory, Los Alamos, NM 87545, USA}
\affiliation{$^5$Department of Mathematics, University of California, Davis, 95616, USA}
\affiliation{$^6$Center for Quantitative Biology, Peking University, Beijing, 100871, China}

\date{\today}

\begin{abstract}
\noindent Line attractors in neuronal networks have been suggested to be the basis of many brain functions, such as working memory, oculomotor control, head movement, locomotion, and sensory processing. In this paper, we make the connection between line attractors and pulse-gating in feedforward neuronal networks. In this context, because of their neutral stability along a one-dimensional manifold, line attractors are associated with a time-translational invariance that allows graded information to be propagated from one neuronal population to the next. To understand how pulse-gating manifests itself in a high-dimensional, non-linear, feedforward integrate-and-fire network, we use a Fokker-Planck approach to analyze system dynamics. We make a connection between pulse-gated propagation in the Fokker-Planck and population-averaged mean-field (firing rate) models, then identify an approximate line attractor in state space as the essential structure underlying graded information propagation. An analysis of the line attractor shows that it consists of three fixed points: a central saddle with an unstable manifold along the line and stable manifolds orthogonal to the line, which is surrounded on either side by stable fixed points. Along the manifold defined by the fixed points, slow dynamics give rise to a ghost. We show that this line attractor arises at a cusp catastrophe, where a fold bifurcation develops as a function of synaptic noise; and that the ghost dynamics near the fold of the cusp underly the robustness of the line attractor. Understanding the dynamical aspects of this cusp catastrophe allows us to show how line attractors can persist in biologically realistic neuronal networks and how the interplay of pulse gating, synaptic coupling and neuronal stochasticity can be used to enable attracting one-dimensional manifolds and thus, dynamically control the processing of graded information.
\end{abstract}


\maketitle


\section{Introduction}
\noindent
Attractor neural networks have been appealed to by theoretical neuroscientists to explain working memory \citep{pmid11476885}, oculomotor control \citep{pmid8917592}, head movement \citep{pmid8604055}, locomotion \citep{pmid25819612}, sensory processing \citep{pmid7731993} and many other experimentally observed brain functions. Some of these neuronal functions have been rationalized using the persistent activity of an attractor (e.g., working memory), while other functions have been modeled by the gradedness (i.e., order preserving property \citep{pmid15718474}) of an attractor (e.g., oculomotor control and sensory processing). Because of these dynamical features, attractor networks have been particularly useful in modeling the encoding of continuous external stimuli \citep{pmid24201281} or the maintenance of internal neuronal representations \citep{pmid15718474}. Furthermore, it has been shown that attractor network dynamics have the capacity to perform optimal computations \citep{pmid15242674} and thus, can implement Bayesian inference \citep{pmid12803954,pmid17522318}.

Pulse-gating is a mechanism capable of transferring packet-based spiking activity from one neuronal population to another \citep{SornborgerWangTao,pmid27310184}. For instance, the synfire-gated synfire chain (SGSC) \citep{pmid27310184} is a mechanism consisting of two sets of neural populations, one representing rate-coded (graded) information and another that gates the flow of that information. This separation into information-carrying and information-control populations provides a basis for understanding how some brain areas can be responsible for the control of information transfer \citep{pmid21267396}, such as the mediodorsal nucleus of the thalamus \citep{pmid12736363}, and others for information processing, such as the somatosensory areas \citep{pmid10811922}.

The original information propagation model that demonstrated the concept of pulse-gated firing rate propagation was based on a mean-field firing rate model of a network of current-based integrate-and-fire (I\&F) neurons that was coarse-grained in time. There was good correspondence between the mean-field model and mean spiking rates in a network of I\&F neurons \citep{SornborgerWangTao,pmid27310184}. However, in general, mean-field models make critical use of the relationship between firing rate and current (f-I curves) (or other input-output functions), which correspond to the steady-state responses of a population of neurons driven by a constant input. There is no {\it a priori} reason that a temporally-averaged mean-field model should correspond with an I\&F network in a transiently driven (pulse-gating) context. Furthermore, the original mean-field solution that demonstrated pulse-gated information propagation suffered from parameter fine-tuning \citep{pmid8917592,Goldman2008,pmid21980334}, even though I\&F simulations appeared robust to parameter variability. 

In this paper, in order to better understand the robustness of pulse-gated, graded information transfer in a network of spiking neurons, we construct Fokker-Planck equations describing the membrane potential probability density function in a feedforward network and study solutions in the state space of a dimensionally-reduced iterative dynamical system. After outlining our methods, we show that pulse-gating in feedforward networks gives rise to approximately time-translationally-invariant spiking probabilities that are propagated from layer to layer in a feedforward network. We then examine synaptic current input-output relations allowing for the construction of an effective population firing rate model and show that this model, averaged across neuronal populations, is very similar to a mean-field model based on rectified linear input-output functions. We show how the input-output function depends on synaptic coupling, gating pulse amplitude, and synaptic noise, and that there is a sizeable region of parameter space within which graded propagation exists.

We then demonstrate the dynamical convergence of membrane potential density to a one-dimensional manifold in parameter space. The Fokker-Planck system reveals that this one-dimensional manifold arises due to a saddle node bifurcation creating an unstable fixed point in the center of a one-dimensional manifold and two stable fixed points on either side of the saddle node. Dynamics orthogonal to the one-dimensional manifold give rise to rapid convergence to the one-dimensional manifold. Ghost (slow) dynamics along the one-dimensional manifold allow the unstable manifold of the saddle to be viewed as an attracting one-dimensional manifold. 

We finally show that this approximate line attractor is robust and generic. By using a reduced analytical model of gating induced transients in the Fokker-Planck system, we show that the propagation of firing amplitudes can be mapped to a cusp catastrophe, and that a bundle of one-dimensional ghost manifolds exists in the region surrounding the fold of the cusp. Our results reveal how the coordination of pulse-gating, synaptic coupling, and membrane potential dynamics enables approximate line attractors in feedforward networks and demonstrate the robustness of graded information propagation when pulse-gating is incorporated.

\section{Methods}
We study a feedforward network of $j = 1, \dots, M$ populations of $i = 1, \dots, N$ excitatory, current-based, integrate-and-fire (I\&F) neurons whose membrane potential, $V$, and synaptic current are described by
\begin{subequations}
\begin{eqnarray}
 \frac{d}{{dt}}V_{i,j}^{} & = &  - {g_L}\left( {V_{i,j}^{} - {V_\mathrm{R}}} \right) + I_{j}^\mathrm{g} + I_{i,j}^{\mathrm{ff}} \label{IFa}\\ 
 \tau \frac{d}{{dt}}I_{i,j}^{\mathrm{ff}} & = &  - I_{i,j}^{\mathrm{ff}} + 
 \left\{ \begin{array}{ll} \frac{S}{{pN}}\sum\limits_k^{} p_{jk}{\sum\limits_l {\delta \left( {t - t_{j-1,k}^l} \right)}}, 
  & j > 1 \\ 
 A \;\delta(t), & j = 1 \end{array} \right. \label{IFb}
\end{eqnarray}
\end{subequations}
where $V_\mathrm{R}$ is the rest voltage (also the reset voltage), $\tau$ is the synaptic timescale, $S$ is the synaptic coupling strength, $p_{jk}$ is a Bernoulli distributed random variable and $p = \langle p_{jk} \rangle$ is the mean synaptic coupling probability. The $l$'th spike time of the $k$'th neuron in layer $j-1$ is determined by $V_{i,j}(t^l_{j-1,k}) = V_\mathrm{Th}$, i.e. when the neuron reaches threshold (after which $V_{i,j}$ is immediately reset to $V_R$). The gating current, $I^g_{j}$, is a white noise process with a square pulse envelope, $\theta(t - (j-1)T) - \theta(t - jT)$, where $\theta$ is a Heaviside theta function and $T$ is the pulse length \citep{SornborgerWangTao} of pulse height $\bar{I}^\mathrm{g}$ and variance $\sigma_0^2$. Note that an exponentially decaying current is injected in population $1$ providing synchronized activity that will subsequently propagate downstream through populations $j = 2, \dots, M$.
\begin{figure}
\centering
  \includegraphics[width=0.45\textwidth]{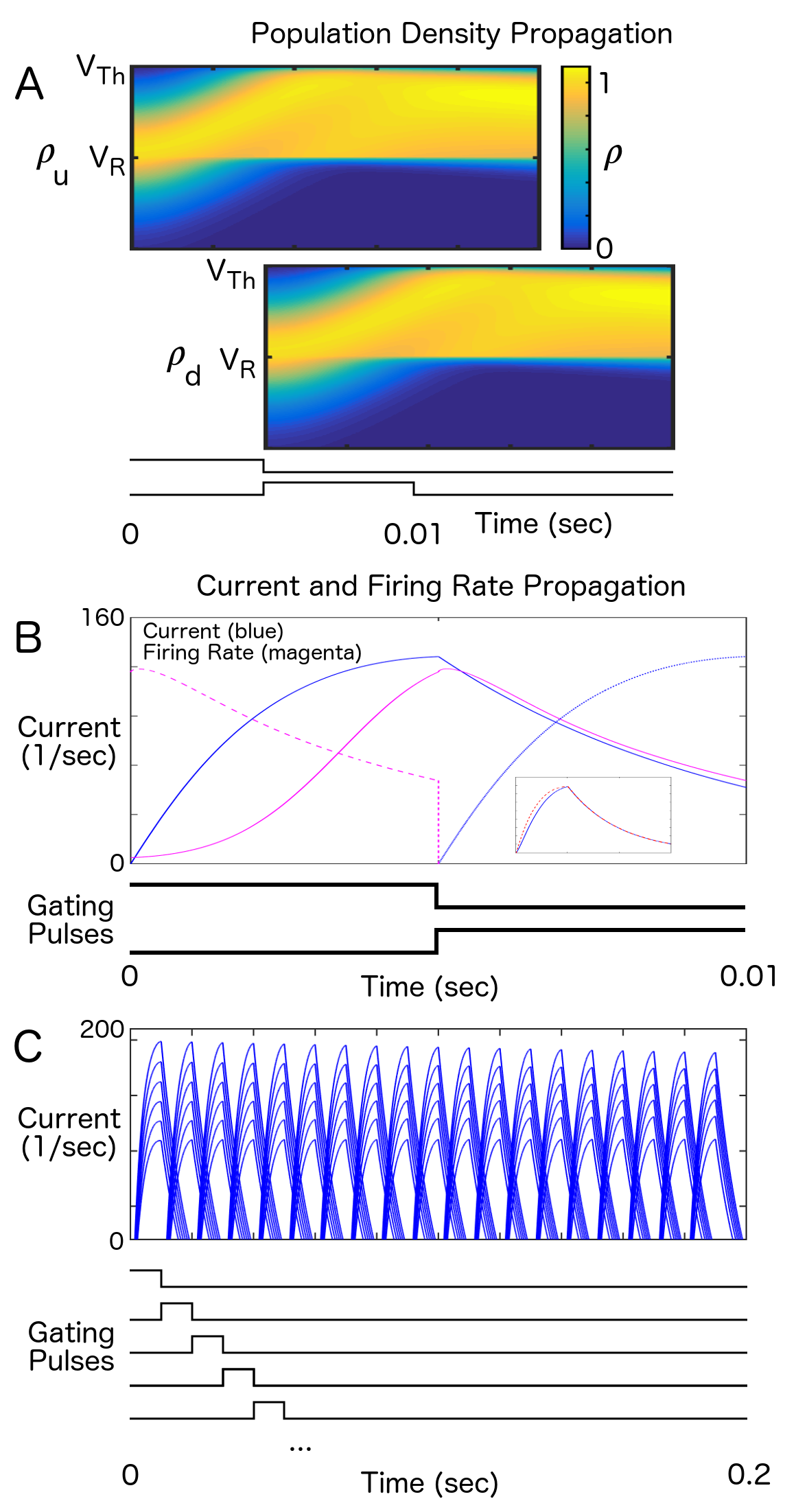}
  \caption{Graded, pulse-gated propagation in a neural Fokker-Planck model. A) Graded population density transfer. Population densities in upstream, $\rho_u(V,t)$, and downstream, $\rho_d(V,t)$, layers. These are two layers from a $19$ layer feedforward network with $T=5$ ms, $S = 2.83$, $I_{gate} = 14.2$, and $\sigma_0 = 4.5$. Pulses gating the upstream and downstream populations are depicted at the bottom of this panel and similarly for subsequent panels. Colorbar applies to both population density plots. B) Synaptic current, $\bar{I}_j^{\mathrm{ff}}$ (blue), and firing rate, $m_j(t)$ (magenta), in upstream and downstream layers. Inset shows a comparison of a mean-field firing rate model (dashed red) with the population-averaged Fokker-Planck model (blue). Below the plot, we show the timing of the gating pulses. C) Graded firing rates showing faithful, pulse-gated propagation across $19$ layers. The sets of successive blue traces (six amplitudes from bottom to top) are currents from each population. Gradedness is shown by the fact that the pulses retain their height (roughly) and order as they propagate from population to population. Gating pulses for the solutions depicted are shown at the bottom of each panel.}
\end{figure}

Feedforward networks of this type admit discrete-time-translationally-invariant solutions describing graded packet transfer  \cite{SornborgerWangTao,pmid27310184}. Such solutions exist for gating pulses that are either temporally sequential or overlapping \cite{pmid27310184}. 

To understand this phenomenon, we used a Fokker-Planck analysis. Assuming the spike trains in Eq. (\ref{IFb}) to be Poisson distributed, the collective behavior of this feedforward network may be described by the (Fokker-Planck) equations
\begin{subequations}\label{FP}
\begin{eqnarray}
 \frac{\partial}{{\partial t}}{\rho _j}\left( {V,t} \right) & = & -\frac{\partial}{\partial V}J_j(V,t)\label{FP1} \\ 
 \tau \frac{d}{{dt}}\overline I _j^{\mathrm{ff}} & = & -\overline I _j^{\mathrm{ff}} + 
  \left\{ \begin{array}{ll} S{m_{j - 1}}, & j > 1 \\
 A\delta(t), & j = 1 \end{array} \right. \label{FP2}
\end{eqnarray}
\end{subequations}
These equations describe the evolution of the probability density function, $\rho_j(V,t)$, in terms of the probability density flux, $J_j(V,t)$, the mean feedforward synaptic current, $\bar{I}^{\mathrm{ff}}_j$, and the population firing rate, $m_j$. For each layer, $j$, the probability density function gives the probability of finding a neuron with membrane potential $V \in (-\infty,V_\mathrm{Th}]$ at time $t$.

The probability density flux is given by
\begin{equation}
J_j\left( {V,t} \right)  =  
 \left( \left[ { - {g_L}\left( {V - {V_\mathrm{R}}} \right) + \bar{I}^\mathrm{g} + \overline I _j^{\mathrm{ff}}} \right] - {D_j}\frac{\partial }{{\partial V}} \right){\rho _j}\left( {V,t} \right)\; \nonumber.
\end{equation}
The effective diffusivity is
\begin{equation}\label{diffusivity}
D_j = \sigma_0^2 + \frac{1}{2}{\frac{{{S^2}}}{{pN}}{m_{j - 1}}\left( t \right)}\;.
\end{equation}
(Below, we take $N \rightarrow \infty$ and neglect the second term on the right in (\ref{diffusivity}).) The population firing rate is the flux of the probability density function at threshold,
\begin{equation}\label{firingrate}
{m_j}\left( t \right) = J_j\left( {{V_\mathrm{Th}},t} \right)\;.
\end{equation}

The boundary conditions for the Fokker-Planck equations are
$J_j\left( {V_\mathrm{R}^ + ,t} \right) = J_j\left( {{V_\mathrm{Th}},t} \right) + J_j\left( {V_\mathrm{R}^ - ,t} \right)$, 
${\rho _j}\left( {V_\mathrm{R}^ + ,t} \right) = {\rho _j}\left( {{V_\mathrm{Th}},t} \right) + {\rho _j}\left( {V_\mathrm{R}^ - ,t} \right)$, 
and
${\rho _j}\left( {V =  - \infty ,t} \right) = 0$.

To improve the efficiency of exploring the bifurcation structure of this system in a large state space, we also investigate an approximate model in which the initial distribution is assumed to be Gaussian, $\rho_0(V,t) = (1/P) \exp{(-(V-\mu(t))^2/2\sigma^2)}$, with width $\sigma$ and mean $\mu(t)$, where $P = \int_{-\infty}^{V_\mathrm{Th}} \rho_0(V,0)$ is a normalization factor accounting for the truncation of the Gaussian at threshold, $V_{\mathrm{Th}}$. As the gating current turns on, the distribution is uniformly advected toward the voltage threshold, $V_\mathrm{Th}$, and the population begins to fire. Uniform advection neglects a small amount of firing due to a diffusive flux across the firing threshold, thus the fold bifurcation (see Results) occurs at a slightly larger value of synaptic coupling, $S$, for this approximation relative to numerical simulations. Since the timescale of the pulse is fast, neurons only have enough time to fire once (approximately). Thus, we neglect the re-emergent population at $V_\mathrm{R}$, which contributes negligibly to firing during the transient pulse.

With this approximation, Eq. (\ref{FP1}) gives rise to $\dot\mu = -g_L (\mu - V_\mathrm{R}) + \bar{I}^\mathrm{g} + \bar{I}_u^{\mathrm{ff}}$, where $\sigma^2 = \sigma_0^2/g_L$. With upstream current $\bar{I}_u^{\mathrm{ff}} = Ae^{-t/\tau}$. Setting $V_{\mathrm{Th}} = 1$, this integrates to
$$\mu(t) = \mu_0 e^{-g_L t} + \frac{\bar{I}^\mathrm{g}}{g_L} (1 - e^{-g_L t}) + \frac{A}{\frac{1}{\tau} - g_L} (e^{-g_L t} - e^{-t/\tau})$$
and from Eq. (\ref{firingrate}), we have
$$m(t) = \left[ (-g_L \mu(t) + \bar{I}^\mathrm{g} + A e^{-t/\tau}) \frac{1}{P} e^{-(1 - \mu(t))^2/2\sigma^2} \right]^+ \;,$$
which, from Eq. (\ref{FP2}), results in a downstream synaptic current at $t = T$
\begin{equation}\label{feedforward}
  \bar{I}_d^{\mathrm{ff}} = S e^{-T/\tau} \int_0^T e^{t/\tau} m(t) \frac{dt}{\tau} \; .
\end{equation}

After the end of the pulse, the current decays exponentially. This decaying current feeds forward and is integrated by the next layer. Thus, for an exact transfer, $\bar{I}_d^{\mathrm{ff}}(S,\bar{I}^\mathrm{g},A,T) = A$.

\begin{figure}
  \centering
  \includegraphics[width=0.5\textwidth]{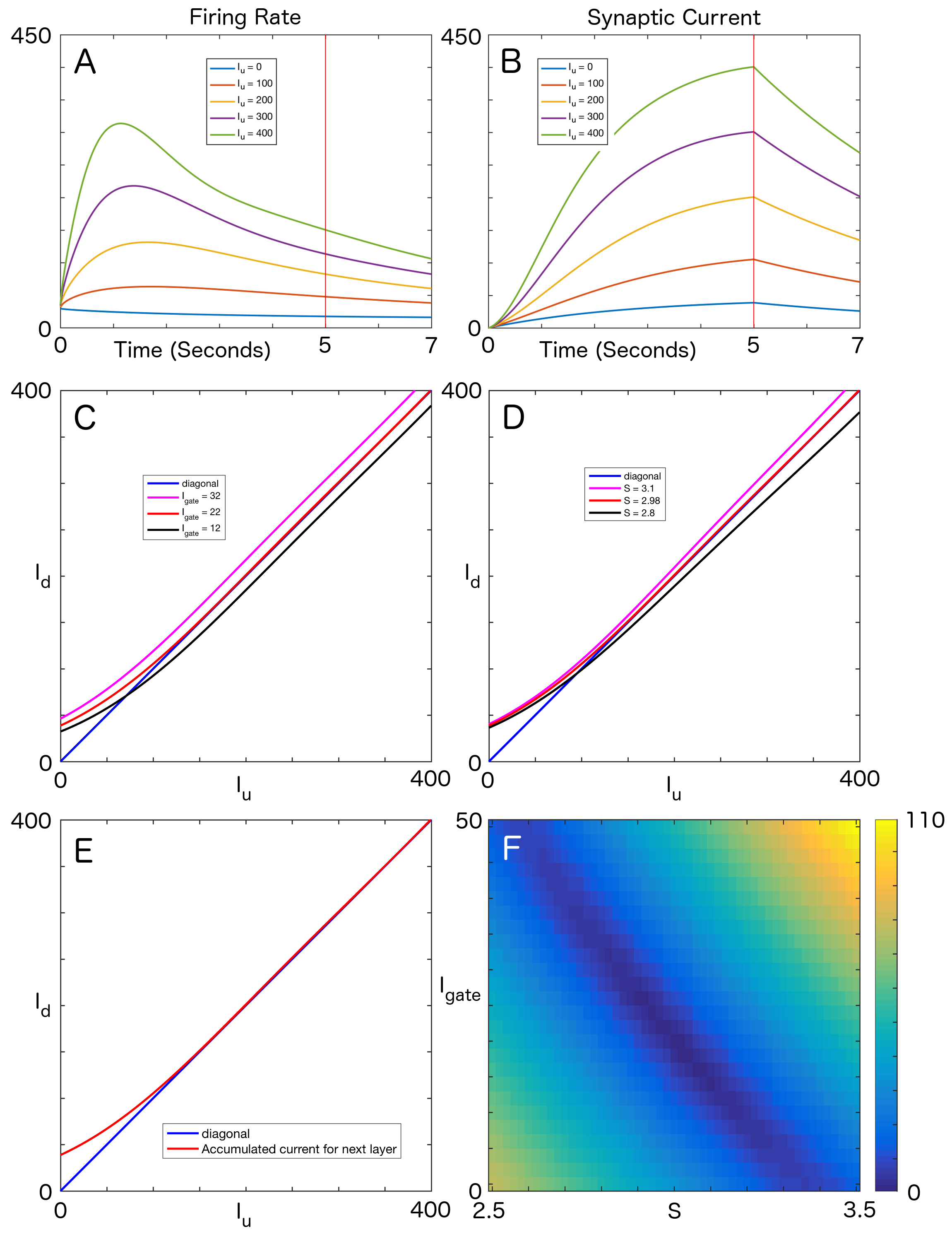}
  \caption{Current transfer in the Fokker-Planck model. A) Downstream firing rates plotted for given upstream synaptic input currents. B) Downstream currents plotted for given upstream synaptic input currents. Note that the peak of the current at the end of the gating pulse, indicated by the vertical red line, is very close to the value of the synaptic input current. For (A) and (B), $\bar{I}^g = 22$,  $T = 5$ ms, $S = 2.98$, and $\sigma = 0.6$. C) Current transfer functions, $I_u(I_d)$, for three values of the gating current, $\bar{I}^g$. Here, $S = 2.98$. Note that changing the gating current moves the input-output function vertically upwards. D) Current transfer functions for three values of the synaptic coupling, $S$. Here, $\bar{I}^g = 22$. Note that changing the synaptic coupling changes the slope of the input-output function. E) An optimal current transfer function, with $S = 2.98$ and $\bar{I}^g = 22$. Note that this transfer function closely approaches the diagonal over a large range of inputs. F) The distance of the current transfer function from the diagonal. The broad, roughly linear, range of gating currents, $\bar{I}^g$, and synaptic couplings, $S$, for which the distance is small indicates a large region of parameter space over which graded information may be faithfully propagated.}
\end{figure}

\section{Results}
In Fig. 1, we show that population density (Fig. 1A), and graded current and firing rate (Fig. 1B) may be propagated via pulse-gating between layers in the Fokker-Planck model. In Fig. 1B inset, we compare currents between a time-averaged firing rate model of pulse-gated graded information propagation and the Fokker-Planck model. For a given value of $T$ in a mean-field rate model, we can find an exact $S$. Because of the slow onset of firing (blue, inset) relative to the firing rate model (dashed red, inset), the value of synaptic coupling that gave rise to graded transfer, $S = S_{graded}$, was larger than the mean-field prediction, $S_{exact}$ (see Appendix), by a factor of $1.07$ for $T$ near $5$ ms. In Fig. 1C, we demonstrate stable graded propagation across many layers.

In Fig. 2A,B, we show firing rates and corresponding current amplitudes for a range of input currents in the Fokker-Planck model. For all but the lowest input current, the output current (the current at the end of the gating pulse) is very close to the input current. Plotting input versus output current for ranges of gating currents, $\bar{I}^g$ (Fig. 2C), and synaptic couplings, $S$ (Fig. 2D), we find that changes in gating currents translate the input-output function upwards and changes in synaptic coupling change the slope of the input-output function. By varying $\bar{I}^g$ and $S$, we can find an optimal input-output function very close to the diagonal (Fig. 2E), hence giving very accurate propagation of graded information. A nearby basin of parameters also gives good graded propagation (Fig. 2F) as measured by the distance of the input-output function from the diagonal. The input-output function in Fig. 2E may be used in a population firing rate model for large pulse-gated systems for which spiking networks are impractical.

In the graded transfer regime, the probability density may be described by a few of its low-lying moments, {\it viz.} $M^1_j(t) = \int_{-\infty}^{V_\mathrm{Th}} V \rho_j(V,t) dV$ and $M^2_j(t) = \int_{-\infty}^{V_\mathrm{Th}} V^2 \rho_j(V,t) dV$. Changes in the moments are related to changes in the shape of the density. Higher moments rapidly converge to the one-dimensional manifold (not shown). To identify time-translationally invariant solutions admitting graded propagation, we search for parameters, $S$, $A$, etc. such that for a given $T$, $\bar{I}_j^{\mathrm{ff}} (t + T) \approx \bar{I}_{j-1}^{\mathrm{ff}}(t)$.

In Fig. 3, we plot the map $(m_j(jT), M^1_j(jT), M^2_j(jT))$ for successive values of $j$ (i.e. mapping feedforward propagation from layer to layer). As may be seen in Fig. 3, for parameters admitting graded information propagation, initial conditions rapidly approach a one-dimensional manifold. On the manifold itself, there are three fixed points, two stable fixed points at the extremes and one unstable saddle in between. For these parameters, the dynamics along the unstable direction of the saddle are slow. On the other hand, the dynamics along the stable direction of the saddle (orthogonal to the one-dimensional manifold) are fast. Thus, after an initial transient taking the packet to the one-dimensional manifold, the amplitude and waveform of each packet remain relatively unchanged from transfer to transfer. Furthermore, the relative ordering of the amplitudes is retained even as the amplitude slowly changes.

\begin{figure}
  \centering
  \includegraphics[width=0.45\textwidth]{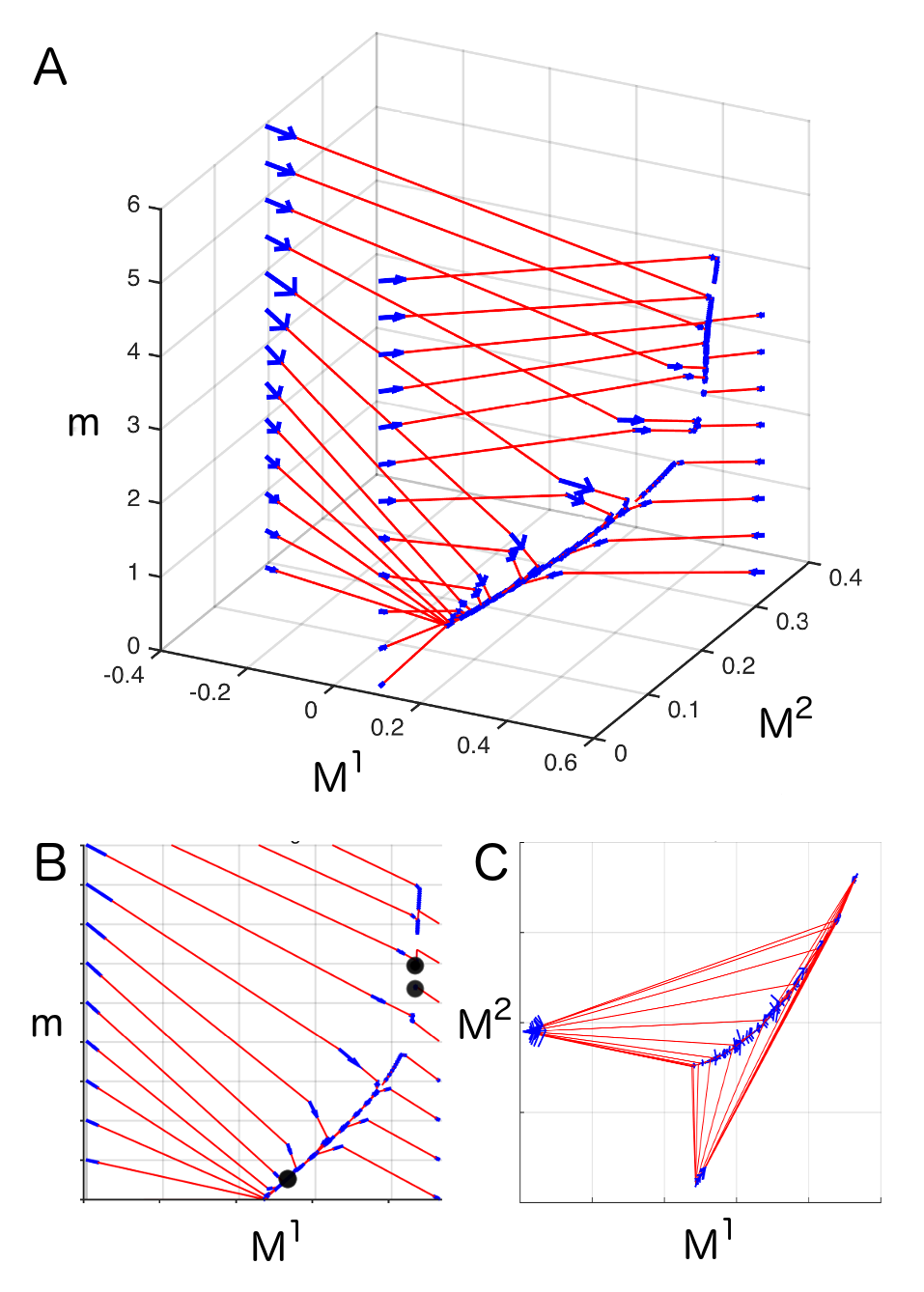}
  \caption{The slow manifold of graded propagation. A) Trajectories, $(m_j(jT), M^1_j(jT), M^2_j(jT))$ for successive values of $j$, showing the rapid approach of the initial distribution to a one-dimensional manifold and the slow evolution along the manifold representing the slow decay of the waveform. The trajectories are plotted as red lines with blue arrows at the tip. Small blue tips along the one-dimensional manifold show the slow evolution. B) The trajectories in A) projected onto the $M^1 - m$ plane. In this panel, the black dots indicate the three fixed points on the one-dimensional manifold (outer fixed points stable, middle fixed point unstable). C) The trajectories in A) projected onto the $M^1 - M^2$ plane. Parameters are as in Fig. 1.}
\end{figure}

Fixed points of the map $(m_j(jT), M^1_j(jT), M^2_j(jT))$ are shown in Fig. 4, showing a fold bifurcation (along the unstable manifold of the saddle) as a function of $S$.

As $S$ moves away from the region of graded transfer, the dynamics along the unstable manifold become fast. Thus, all amplitudes rapidly approach the fixed points at the extremes. This leads to a mechanism for binary information transfer (in distinction from graded information transfer), where pulse gating can only transfer low and high amplitudes. As $S$ increases, the location of the high fixed point increases giving more distinction between low and high fixed points.

This behavior may be understood across a large region of parameter space with the use of the approximate, Gaussian model (see Methods).

\begin{figure}
  \centering
  \includegraphics[width=0.4\textwidth]{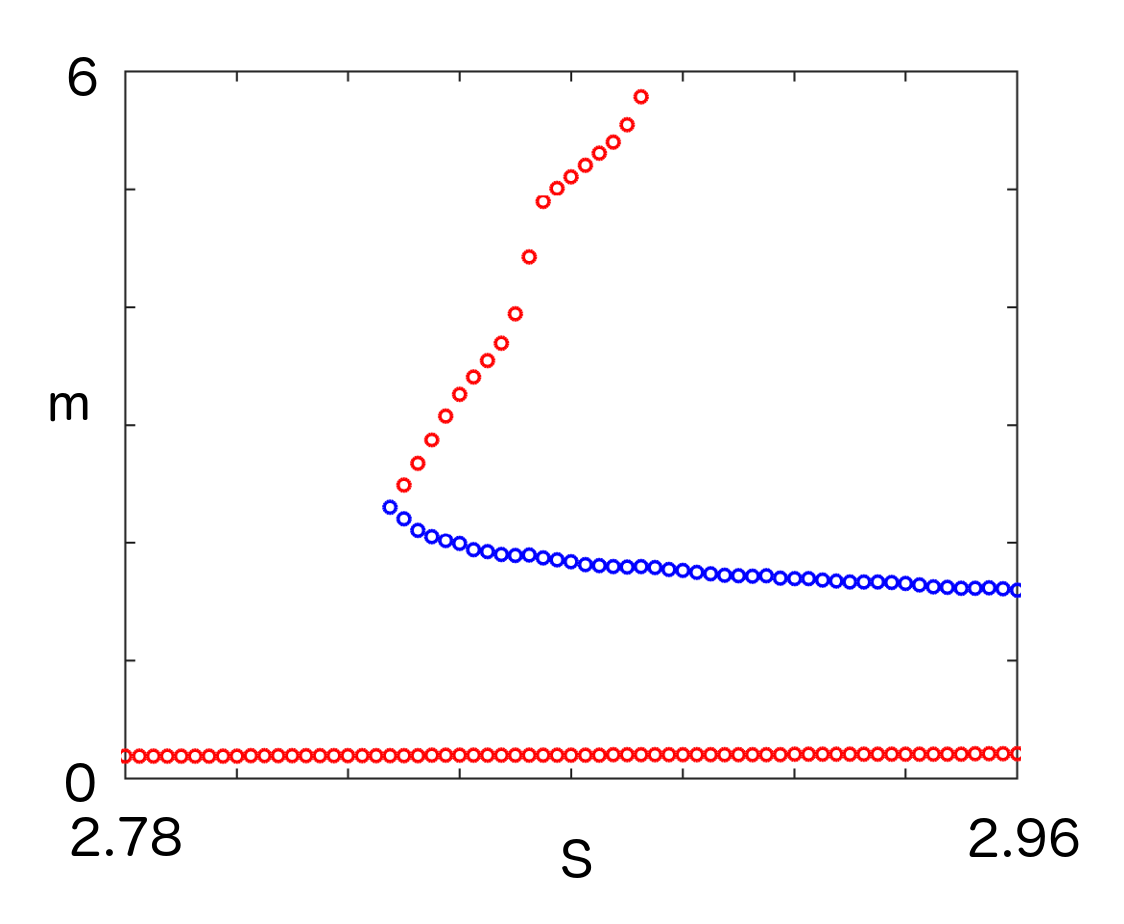}
  \caption{Bifurcation diagram generated using numerically integrated Fokker-Planck equations (\ref{FP}), for $I = 14.2$. The saddle-node bifurcation occurs at approximately $S = 2.82$. Blue indicate unstable fixed points, and red indicates stable fixed points. For values close to $S = 2.82$ the trajectory evolves slowly along the one-dimensional manifold due to a fold bifurcation ghost, allowing graded propagation across many layers before significant change in the waveform.}
\end{figure}

In Fig. 5A,B, we compare numerical and approximate (Gaussian) solutions. Note the similar advection of the population density with obvious differences in curvature and reset populations (which may be seen in Fig. 5A as a linear wall of density along the reset value, $0$, of the membrane potential). In Fig. 5C, we show the fold bifurcation, found numerically in Fig. 4, for a range of values of $S$ and $\bar{I}^\mathrm{g}$. In Fig. 5D, we demonstrate the existence of a cusp catastrophe as a function of the width of the potential distribution $\sigma$. In Fig. 5E,F, we show how fast or slow (ghost) dynamics occur near the fold of the cusp depending on $\sigma$ and $S$. In Fig. 5E,F, blue shades indicate locations where feedforward amplitudes are slowly changing, green and yellow shades where amplitudes change rapidly. Isoclines (the same in both panels) are plotted denoting the fixed point (central isocline) and nearby locations of small, but non-zero change. Note how in Fig. 5E,F, dynamics are fast and amplitudes rapidly approach the attractors for large synaptic coupling, but dynamics remain slow in the region near the fold bifurcation. This region gives rise to an approximate line attractor that becomes more robust, {\it i.e.} covering a large range of amplitudes ($A/g_L$), as $\sigma$ increases (compare Fig. 5E and Fig. 5F).

\begin{figure}
  \includegraphics[width=0.5\textwidth]{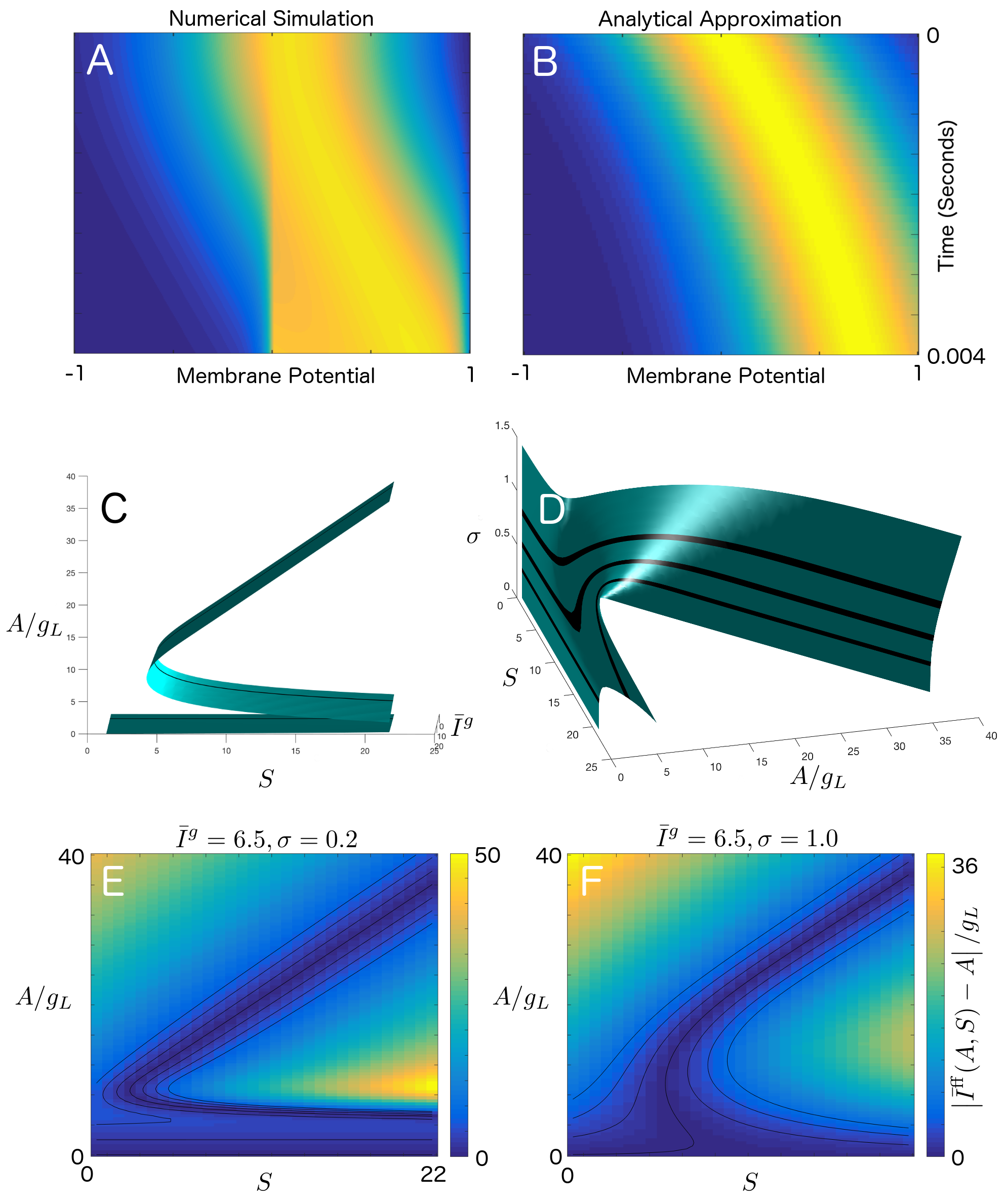}
  \caption{$\sigma$-induced fold bifurcation ghost at cusp catastrophe. A) The probability density as a function of time and membrane potential from a numerical simulation. B) The probability density as a function of time from the transient pulse approximation. C) A plot of the surface of zeros of $\bar{I}^{\mathrm{ff}}(S,\bar{I}^\mathrm{g},A,\sigma) - A$ plotted for $(S,\bar{I}^\mathrm{g},A)$ with $\sigma = 0.2$. These are fixed points of the map of the synaptic feed forward current generated from Eqs. (\ref{feedforward}). Note the fold bifurcation across many values of $\bar{I}^\mathrm{g}$, similar to that found numerically in Fig. 3. D) A plot of the surface of zeros of $\bar{I}^{\mathrm{ff}}(S,\bar{I}^\mathrm{g},A,\sigma) - A$ plotted for $(S,A,\sigma)$ with $\bar{I}^\mathrm{g} = 6.5$. Note the obvious cusp catastrophe that occurs along the $\sigma$-axis. E) Zeros of $\bar{I}^{\mathrm{ff}}(S,\bar{I}^\mathrm{g},A,\sigma) - A$ for $\bar{I}^\mathrm{g} = 6.5$ and $\sigma = 0.2$ plotted on a colored background, where the color indicates the speed of departure toward stable fixed points (either upward above the unstable fixed point or downward below the unstable fixed point). For small $\sigma$, solutions rapidly approach the fixed points, giving a propagated binary code for large $S$. F) Zeros of $\bar{I}^{\mathrm{ff}}(S,\bar{I}^\mathrm{g},A,\sigma) - A$ for $\bar{I}^\mathrm{g} = 6.5$ and $\sigma = 1.0$ again indicating the speed of departure toward stable fixed points. For $\sigma$ near the cusp, there is a large region within which propagation away from the unstable point is slow, called a ghost. Ghosts occur near the cusp locus producing a bundle of models in parameter space that allow graded propagation across many layers.}
\end{figure}

\section{Discussion}
To understand pulse-gated information propagation, we used a Fokker-Planck analysis to derive input-output curves that may be used in population firing rate models of pulse-gated propagation and to reveal an approximate line attractor in the network of spiking neurons. We further showed that the line attractor is associated with ghost dynamics occurring along the front fold (relatively small synaptic coupling) of a cusp catastrophe as $\sigma$ varies in state space. Locally, the nearly one-dimensional attracting manifold is the result of a fold bifurcation where the stable directions of the saddle and attracting fixed points are strongly attracting and the unstable manifold exhibits ghost dynamics. In terms of input-output functions, this reflects that the fold of this cusp separates a region of sigmoidal f-I curves, with a relatively large linear interval, from a region of bistability. Since this region occupies a sizeable volume in parameter space, the graded propagation is robust. Furthermore, the fact that a common cusp catastrophe underlies the fast timescale dynamics of our system indicates that this type of line attractor is generic and will persist in feedforward networks of other types of spiking neurons.

One of the major problems of neuroscience is to understand how complex neural functions emerge from the collective dynamics of neuronal networks. Towards this goal, researchers have tried to construct models using mean-field firing rate theories and large-scale numerical simulations. However, except for a few examples (e.g. \citep{pmid9962697,NykampTranchina2000,CaiKinetic2006,pmid21283777,MontbrioEtAl2015,pmid27494737}), the precise correspondence between the underlying microscopic spiking neurons and the macroscopic coherent dynamics of the neuronal populations has not been established.

Our analysis provides a significant step toward understanding how macroscopic attracting manifolds can emerge from the dynamic interactions of  microscopic spiking neurons via the coordination of pulse-gating, synaptic weights, and intrinsic and extrinsic noise (i.e. the distribution of the membrane potential across the population), and offers possible order parameters for which macroscopic descriptions can be derived from the underlying microscopic dynamical model. 

Furthermore, transfer mechanisms, such as those found in the graded and binary transfer parameter regimes shown in Fig. 5. ({\it i.e.} for $S$ near the fold (graded) or $S$ to the right of the fold (binary)), provide a novel means of understanding dynamic network interactions such as the detailed measurements of population activity underlying complex neural tasks provided by modern experimental techniques. Already, pulse-gated transfer mechanisms have been shown to be capable of implementing dynamic modules representing complex neural functions, such as short term memory, decision making, and the control of neural circuits \citep{SornborgerWangTao,pmid27310184}.

\section{Appendix: Mean Field Model}

A mean-field firing rate model of Eq. (1) is given by
\begin{eqnarray}
  \tau \frac{d\bar{I}_j}{dt}^{ff} & = & -\bar{I}^{ff}_j + Sm_{j-1} \\
  m_j & = & \left[ \bar{I}^{ff}_j + I^g_j(t) - g_0 \right]^+
\end{eqnarray}
where $[ ]^+$ denotes a thresholded linear function, with threshold $g_0$, for the input-output relation of an I\&F neuron.

In \citep{SornborgerWangTao}, we showed that when the gating pulse cancels the threshold ($\bar{I}^g = g_0$), and the feedforward synaptic coupling strength was
\begin{equation}
  S_{exact} = \frac{\tau}{T}e^{T/\tau}
\end{equation}
we get exact, graded propagation, where the mean synaptic current and firing rates were
\begin{equation}
  I^{ff}_j(t) = \left\{ \begin{array}{ll}
           A\left( \frac{t - (j - 1)T}{\tau} \right) e^{-(t - (j-1)T)/\tau}, & (j-1)T \le t \le jT \\
           A \left( \frac{T}{\tau} e^{-T/\tau} \right) e^{-(t - jT)/\tau}, & jT < t < \infty \end{array} \right.
\end{equation}
and
\begin{equation}
  m_j(t) = \left\{ \begin{array}{ll}
           0, & 0 < t <(j-1)T \\
           A \left( \frac{T}{\tau} e^{-T/\tau} \right) e^{-(t - jT)/\tau}, & (j-1)T \le t \le jT \\
           0, & jT < t < \infty \end{array} \right.
\end{equation}

\begin{acknowledgments}
L.T. thanks the UC Davis Mathematics Department for its hospitality. A.T.S. would like to thank Liping Wei and the Center for Bioinformatics at the School of Life Sciences at Peking University for their hospitality. This work was supported by the Natural Science Foundation of China grants 91232715 (Z.X, L.T.), 31771147 (L.T.), 91430216 (J.Z.), and U1530401 (J.Z.), by the Open Research Fund of the State Key Laboratory of Cognitive Neuroscience and Learning grant CNLZD1404 (Z.X., L.T.), by the Beijing Municipal Science and Technology Commission under contract Z151100000915070 (Z.X., L.T.), and by the Undergraduate Honors Research Program of the School of Life Sciences at Peking University (Z.X., L.T.).
\end{acknowledgments}

\bibliographystyle{unsrtnat}
\bibliography{Biblio}

\end{document}